\documentclass{article}

    \PassOptionsToPackage{numbers, compress}{natbib}

\usepackage[final]{tackling_climate_workshop_style}

\usepackage{natbib} 
\usepackage[utf8]{inputenc} 
\usepackage[T1]{fontenc}    
\usepackage{hyperref}       
\usepackage{graphicx}
\usepackage{url}            
\usepackage{booktabs}       
\usepackage{amsfonts}       
\usepackage{nicefrac}       
\usepackage{microtype}      
\usepackage{nicefrac}       
\usepackage{graphicx}
\usepackage{overpic}
\usepackage{comment}
\usepackage{array}
\usepackage{amsmath}

\title{\textsc{Pyrocast}: a Machine Learning Pipeline to Forecast Pyrocumulonimbus (PyroCb) Clouds}

\author{Kenza Tazi\\ 
        Department of Engineering\\
        University of Cambridge\\
        Cambridge, UK\\
        \texttt{kt484@cam.ac.uk}\\
        \And
        Emiliano Díaz Salas-Porras\\
        Department of Electrical Engineering\\
        University of Valencia\\
        Valencia, Spain\\
        \texttt{emiliano.diaz@uv.es}\\
        \And
        Ashwin Braude\\
        Laboratoire Atmosphères,\\
        Milieux, Observations Spatiales\\
        Institut Pierre-Simon Laplace\\
        Guyancourt, France\\
        \And
        Daniel Okoh\\
        Centre for Atmospheric Research\\
        National Space Research \\
        and Development Agency\\
        Abuja, Nigeria\\
        \And
        Kara D. Lamb\\
        Department of Earth and \\
        Environmental Engineering\\
        Columbia University\\
        New York, US\\
        \And
        Duncan Watson-Parris\\
        Department of Atmospheric,\\ 
        Oceanic and Planetary Physics\\
        University of Oxford\\
        Oxford, UK\\
        \And
        Paula Harder\\
        Fraunhofer Institute for \\
        Industrial Mathematics\\
        Kaiserslautern, Germany\\
        \And
        Nis Meinert\\
        Pasteur Labs \& ISI\\
        New York, US\\
        }

\begin{document}

\maketitle

\begin{abstract}
Pyrocumulonimbus (pyroCb) clouds are storm clouds generated by extreme wildfires. PyroCbs are associated with unpredictable, and therefore dangerous, wildfire spread. They can also inject smoke particles and trace gases into the upper troposphere and lower stratosphere, affecting the Earth's climate. As global temperatures increase, these previously rare events are becoming more common. Being able to predict which fires are likely to generate pyroCb is therefore key to climate adaptation in wildfire-prone areas. This paper introduces \textsc{Pyrocast}, a pipeline for pyroCb analysis and forecasting. The pipeline's first two components, a pyroCb database and a pyroCb forecast model, are presented. The database brings together geostationary imagery and environmental data for over 148 pyroCb events across North America, Australia, and Russia between 2018 and 2022. Random Forests, Convolutional Neural Networks (CNNs), and CNNs pretrained with Auto-Encoders were tested to predict the generation of pyroCb for a given fire six hours in advance. The best model predicted pyroCb with an AUC of $0.90 \pm 0.04$.
\end{abstract}

\section{Introduction}

More than 17 million people have been affected and USD\,144 billion lost through major wildfire events over the last 30 years \citep{guha2015}. In addition, the degradation of air quality due to the creation of aerosols and ozone from fires results in between 260\,000 and 600\,000 premature deaths each year \citep{johnston2012estimated}. The risk posed by wildfires to people and the environment is increasing due to climate change. By the end of the century, the frequency of wildfires, compared to a 2000 -- 2010 reference period, is predicted to increase by a factor of $1.31$ to $1.57$ with the number of extreme wildfires increasing even further \citep{unep2022, dowdy2019future}.

Pyrocumulonimbus (pyroCb) clouds are produced by large and intense wildfires \citep{peterson2017detection, rodriguez2020extreme}. PyroCbs create their own weather fronts which can make wildfire behaviour unpredictable through strong winds and ignite new fires through lightning strikes \citep{peterson2021australia}. PyroCbs also inject wildfire aerosols and trace gases into the stratosphere where they remain for several months \citep{peterson2021australia, yu2019black, peterson2018wildfire}. These events, which can be on the scale of a volcanic eruption, have important impacts on the Earth's climate, e.g., through direct absorption of solar radiation or cloud formation \citep{ke2021global, stocker2021observing, kablick2020australian}. PyroCbs could also hinder the recovery of the ozone layer \citep{rieger2021stratospheric, schwartz2020australian}.

Despite the risk posed by pyroCbs, the conditions leading to their occurrence and evolution are still poorly understood. Previous pyroCb research has generally been limited to the study of occurrences linked to single wildfire events or over a limited study areas. Geostationary and local weather information has been used to characterise the properties of these clouds \citep{sienko, peterson2014quantifying} and to create an empirical thresholding algorithm to detect pyroCb from GOES17 satellite imagery \citep{peterson2017detection}. This paper presents a machine learning pipeline named ‘\textsc{Pyrocast}’, with the aim of monitoring, forecasting, and understanding the drivers behind pyroCb events. This pipeline will aid scientists to systematically research pyroCb, as well as help policymakers and emergency services efficiently allocate resources and evacuate residents and emergency responders.

\textsc{Pyrocast} \footnote{Code and data are available at \url{https://doi.org/10.56272/fpib2524}} is made up of three components: 
\begin{enumerate}
 \item A database (Section \ref{section:data}) containing labelled geostationary satellite images, meteorological, and fuel data associated with individual wildfire events. To the best of the authors' knowledge, the database is the most comprehensive pyroCb database published to date.
 \item A forecast model (Section \ref{section:model}), a Random Forest (RF) model that can predict pyroCb occurrence from a wildfire over a six-hour horizon. This is the first model that attempts to forecast pyroCb and the first application of machine learning to this field. Convolutional Neural Networks (CNNs) and CNNs pretrained with Auto-Encoders (AE-CNNs) are also explored in this study.
 \item A causal discovery framework, a set of tools to understand the characteristics and causes of pyroCb published in a companion paper \citep{pyrocast_causality}. This is the first attempt to create a causal model for pyroCb formation.
\end{enumerate}

\section{\textsc{Pyrocast} Database}
\label{section:data}

To create the \textsc{Pyrocast} database, historical pyroCb events were manually collated from blogs and sparse databases including the Australian PyroCb registry \citep{aus_pyrocb}, the CIMSS PyroCb Blog \citep{cimss_pyrocb}, Annastasia Sienko's master's thesis \citep{sienko} and the PyroCb Online Forum \citep{pyrocb_forum}. The events were then matched to known historical wildfire events in the GlobFire database \citep{artes2019global} to determine the start and end date of each fire. PyroCb that could not be associated with any wildfire were given arbitrary star and end dates, three days before and after the pyroCb sighting respectively. 

Geostationary satellite imagery from Himawari-8, GOES16, and GOES17 was then matched to the wildfire locations and dates. The spatial coverage of these satellites overlaps with most of the recorded pyroCb located in North America, Australia and Russia. The satellite imaging instruments also have high temporal (10\,min) and spatial resolutions ($0.5-2$\,km) to study the evolution of the wildfires. Six wavelength channels ($0.47$\,\textmu m, $0.64$\,\textmu m, $0.86$\,\textmu m, $3.9$\,\textmu m, $11.2$\,\textmu m, and $13.3$\,\textmu m) were chosen to detect and predict pyroCb. Images were downloaded on an hourly basis during local daytime hours at the wildfire locations from Amazon Web Services \citep{aws_goes, aws_himawari} using a custom parallelisation pipeline. Each scene was interpolated to 1km resolution and cropped to 200 by 200-pixel image.

The cropped geostationary images were then fed into a pyroCb detection algorithm developed by Peterson et al.\ at the US Naval Research Laboratory (NRL) \citep{peterson2017detection}. The algorithm was used to output pyroCb masks and flags for the forecast model. Finally, the geostationary images were matched with meteorological and fuel data from a climate reanalysis model. For this study, ERA5 reanalysis \citep{hersbach2020era5} was used as it is global, up-to-date, accurate, has a high temporal resolution (hourly), and is easily accessible via the Copernicus Data Store API \citep{cdsapi}. Nineteen variables were downloaded and interpolated to the geostationary grid. The full list of geostationary satellite and ERA5 variables, with reasons for selection, can be found in Table \ref{intext_variabletable} of the \nameref{appendix}. Overall, the \textsc{Pyrocast} database includes over 148 PyroCb events linked to 111 wildfires between 2018 and 2022, equivalent to over 18 thousand hourly observations. Most importantly, it is science and machine learning ready, allowing for the systematic study of the characteristics and causes of PyroCb.

\section{\textsc{Pyrocast} Forecast Model}
\label{section:model}

\subsection{Setup}
To create a pyroCb forecast model, multiple experiments with a combination of model architectures were performed: Random Forest (RF), Convolutional Neural Network (CNN), and Auto-Encoder pretrained Convolutional Neural Network (AE-CNN). \texttt{PyTorch} was used to implement CNNs and AE, and \texttt{Scikit Learn} for the RFs and metrics. Models were trained for three different learning tasks: `detection' where labels and input correspond to the same timestamp, `forecast with weather oracle' where output labels and meteorological inputs correspond to timestamps six hours after that of the geostationary inputs, and `forecast' where output labels correspond to timestamps six hours after that of the geostationary and meteorological inputs.
 
The models were trained on five different types of input: geostationary channels (gs), three meteorological variables (w3) selected by importance using a RF for a detection task (convective available potential energy, boundary layer height, and relative humidity at $650$\,hPa), geostationary channels and important meteorological variables (gs+w3), all meteorological variables (w19), and geostationary and all meteorological variables available (gs+w19). Only data from North America and Australia were used for training. A cross-validation scheme with five folds was used given that multiple pyroCb observations can belong to the same wildfire event. The folds were defined by randomly assigning wildfire events to one of five clusters. A fold scheme defined by clustering by latitude and longitude is also presented in the \nameref{appendix}. These results correspond to the performance that could be expected applying the forecast model to an unseen region such as Europe.

\subsection{Training}
The RF models consisted of 500 trees, where tree splits were chosen to minimise Gini impurity. For the RF models, 11 features were aggregated across the spatial dimensions: mean, standard-deviation, minimum, maximum, and the quantiles corresponding to the following percentiles: $0.01$, $0.05$, $0.25$, $0.5$, $0.75$, $0.95$, $0.99$. Trees were grown to a maximum depth of 10 splits. Each tree was trained on a bootstrap subsample of the data. In evaluating the quality of splits, weighting was used to compensate for unbalanced labels.

The CNN model consisted of six convolutional layers, followed by two fully connected layers. Max pooling was used after each convolutional layer and drop-out (25\% drop-out probability) was used after the convolutional layers. ReLU activation functions were used for each layer. A cross-entropy loss function was used to train the CNN classifier. ADAM optimization with a batch size of 64 and a learning rate of $0.001$ was performed to train the model. CNN models were trained for five epochs only as overfitting was observed after a low number of epochs. This is due to the small amount of independent data (84 events, and $\sim$6k training observations).

AEs were used to obtain pretrained initializations for the weights of the CNN. The CNN described above corresponds to the encoder and a mirror decoder was used to project the final 16-dimensional hidden layer to outputs with the same dimension as the inputs. The mean squared error (MSE) was used to measure the reconstruction loss. We also used ADAM optimization with a batch size of 64 and learning rate of $0.001$ to train the AEs. Given the richer labels, the AE-CNN models were trained for 40 epochs without overfitting.

\subsection{Results}
 The average `Area Under the Curve' (AUC) for each experiment is shown in Table \ref{auc_avg}. For the `detection' task, the geostationary channels contributed significantly more to the performance. The AE-CNN and RF trained on the geostationary channels were the best detection models. The performance of the `forecast with weather oracle' and `forecast' models are similar. In practice, however, the performance for the `forecast with weather oracle' will diminish when the oracle is replaced by a weather forecast.

\begin{table}[t]
\caption{Average validation AUC across 5 folds with standard deviations.}
\label{auc_avg}
\centering
\small
\begin{tabular}{llccccc}
\toprule
 & & gs & w3 & gs+w3 & w19 & gs+w19\\

\midrule
Detection & RF & $\mathbf{0.96 \pm 0.00}$ & $0.84 \pm 0.04$ & $0.96 \pm 0.01$ & $0.90 \pm 0.04$ & $0.96 \pm 0.01$\\
& CNN & $0.88 \pm 0.19$ & $0.69 \pm 0.06$ & $0.89 \pm 0.14$ & & \\
& AE-CNN & $\mathbf{0.97 \pm 0.01}$ & $0.72 \pm 0.05$ & $0.94 \pm 0.01$ & & \\

\midrule
Forecast & RF & $0.84 \pm 0.03$ & $0.83 \pm 0.06$ & $0.88 \pm 0.04$ & $0.89 \pm 0.05$ & $\mathbf{0.90 \pm 0.04}$ \\
with oracle & CNN & $0.55 \pm 0.06$ & $0.67 \pm 0.06$ & $0.66 \pm 0.10$ &  & \\
& AE-CNN & $0.62 \pm 0.06$ & $0.72 \pm 0.06$ & $0.72 \pm 0.04$ & & \\

\midrule
Forecast & RF & $0.84 \pm 0.03$ & $0.82 \pm 0.05$ & $0.87 \pm 0.03$ & $0.88 \pm 0.05$ & $\mathbf{0.90 \pm 0.04}$\\
& CNN & $0.55 \pm 0.06$ & $0.73 \pm 0.03$ & $0.71 \pm 0.05$ &  & \\
& AE-CNN & $0.62 \pm 0.06$ & $0.72 \pm 0.03$ & $0.75 \pm 0.05$ &  & \\
\bottomrule
\end{tabular}
\end{table}

 For the forecast tasks, both the geostationary and meteorological data contributed to the predictive power of the models. This agrees with what the literature has suggested so far: it takes a combination of a very strong wildfire (which the geostationary imagery can detect) and the right meteorological conditions to generate a pyroCb event \citep{peterson2017conceptual, tory2021pyrocumulonimbus, tory2018thermodynamics}. The RF achieved the best performance: $0.90 \pm 0.04$ AUC for both the `forecast with weather oracle' and `forecast' tasks using all the available variables. The quantity of data used to train the models may be too small to leverage the full potential of a CNN model and to exploit the spatially resolved information in the geostationary and meteorological data. The 11 aggregated features may be enough to extract most of the relevant information to predict pyroCb.

\begin{figure}[h]
    \centering
        \includegraphics[scale=0.232]{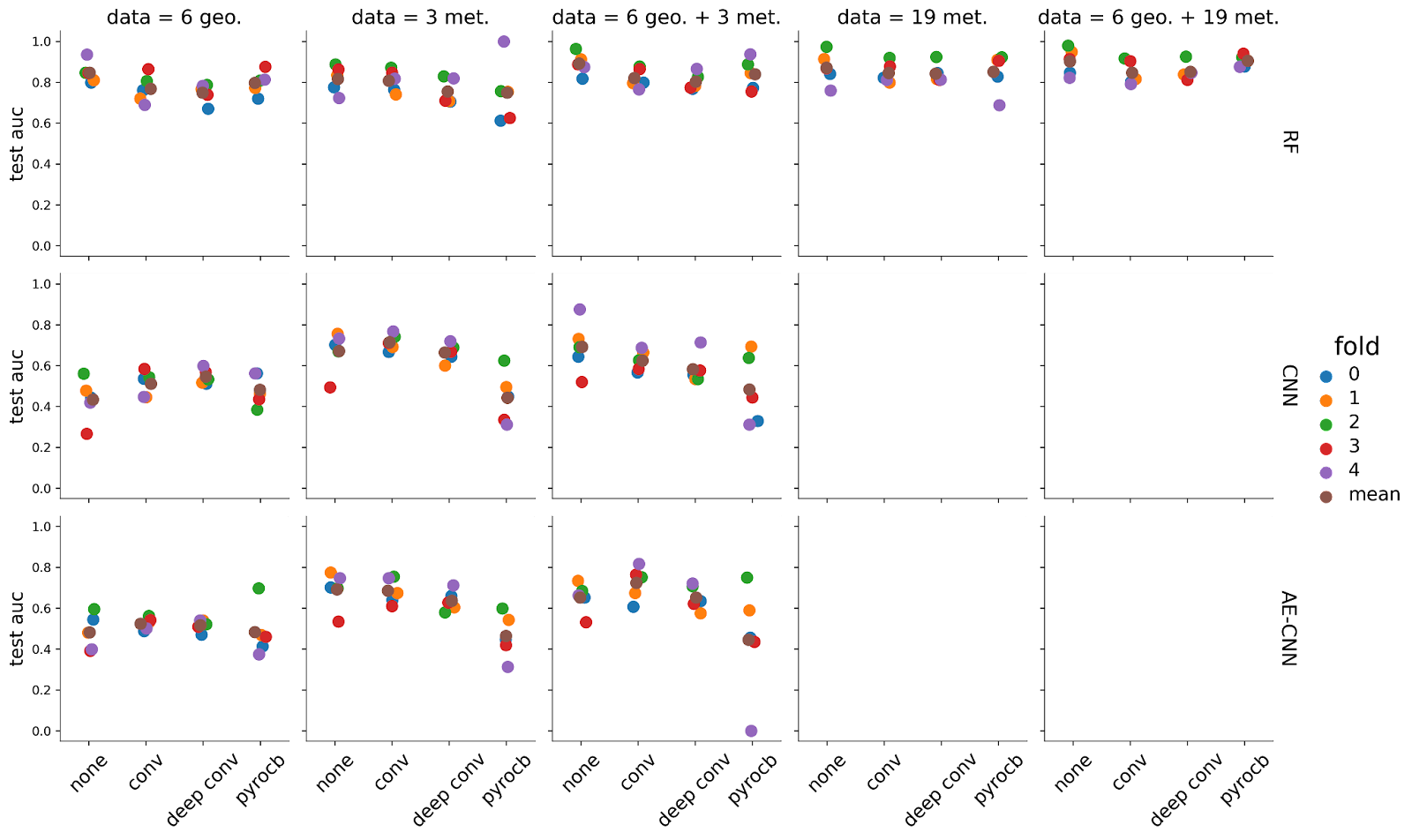}
        \caption{Validation AUC for different forecast models and input sets as a function of the NRL flag: `None' for no convection or pyroCb observed, `Convection' for when convection is observed but no deep convection or pyroCb, `Deep-convection' for when deep convection is observed but no pyroCb, `PyroCb' for when a pyroCb is already detected.}
        \label{fig:wildfire_state}
\end{figure}

The performance of the `forecast' model as a function of the wildfire state was also analysed. These states correspond to a grouping of the NRL algorithm flags (cf.\ Section \ref{section:data}). Figure \ref{fig:wildfire_state} summarises the AUC of the forecast model as a function of the current state of the wildfire. In general, the RF model performed best when there was no sign of convection or pyroCb six hours before the label was recorded. On the other hand, models struggled the most to predict the pyroCb when pyroconvection was already detected.

\section{Conclusion}

This paper presents \textsc{Pyrocast}, a pipeline to monitor, forecast, and understand the drivers behind pyroCbs clouds formed by extreme wildfires. The \textsc{Pyrocast} database provides the most comprehensive pyroCb dataset published to date with labelled geostationary satellite images and environmental data for 148 pyroCb events across North America, Australia and Russia between 2018 and 2022. The \textsc{Pyrocast} forecast model is trained using this data. A RF model showed the most skill in forecasting pyroCb occurrence from a wildfire six hours in advance with an AUC of $0.90 \pm 0.04$. This is the first attempt to forecast pyroCb and the first application of machine learning to this field. Further work could include expanding the database through the creation and application of a detection model for PyroCb and PyroCu clouds, the precursors of PyroCb. More data would in turn improve forecast performance. Finally, an evaluation of model uncertainty and confidence could also enable better decision making if the forecast model is deployed during a wildfire.

\newpage
\section*{Acknowledgements}
This work is the result of the 2022 Frontier Development Lab Europe Research Sprint. We are grateful for the support of the organizers, mentors, and sponsors. In particular, we would like to thank David Peterson, Michael Fromm, Annastasia Sienko, and Raul Ramos for their insight and help. Funding for this study was also provided by the European Research Council (ERC) Synergy Grant "Understanding and Modelling the Earth System with Machine Learning (USMILE)” under the Horizon 2020 research and innovation programme (Grant agreement No. 855187).

\bibliography{main.bib}

\newpage
\section*{Appendix}
\label{appendix}

\begin{figure}[h]
    \centering
        \includegraphics[scale=0.24]{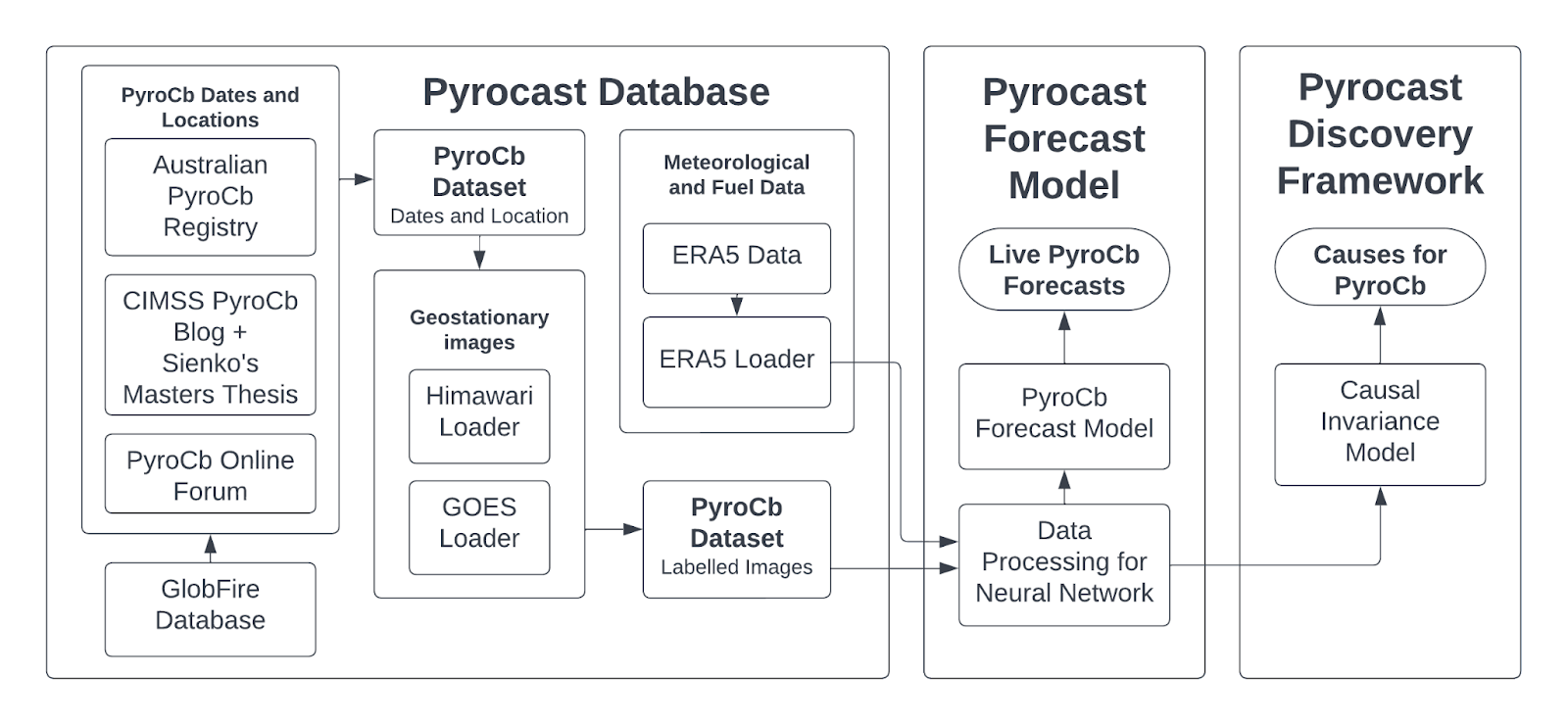}
        \caption{Diagram of \textsc{Pyrocast} pipeline.}
        \label{fig:pipeline}
\end{figure}

\begin{figure}[h]
    \centering
        \includegraphics[scale=0.4]{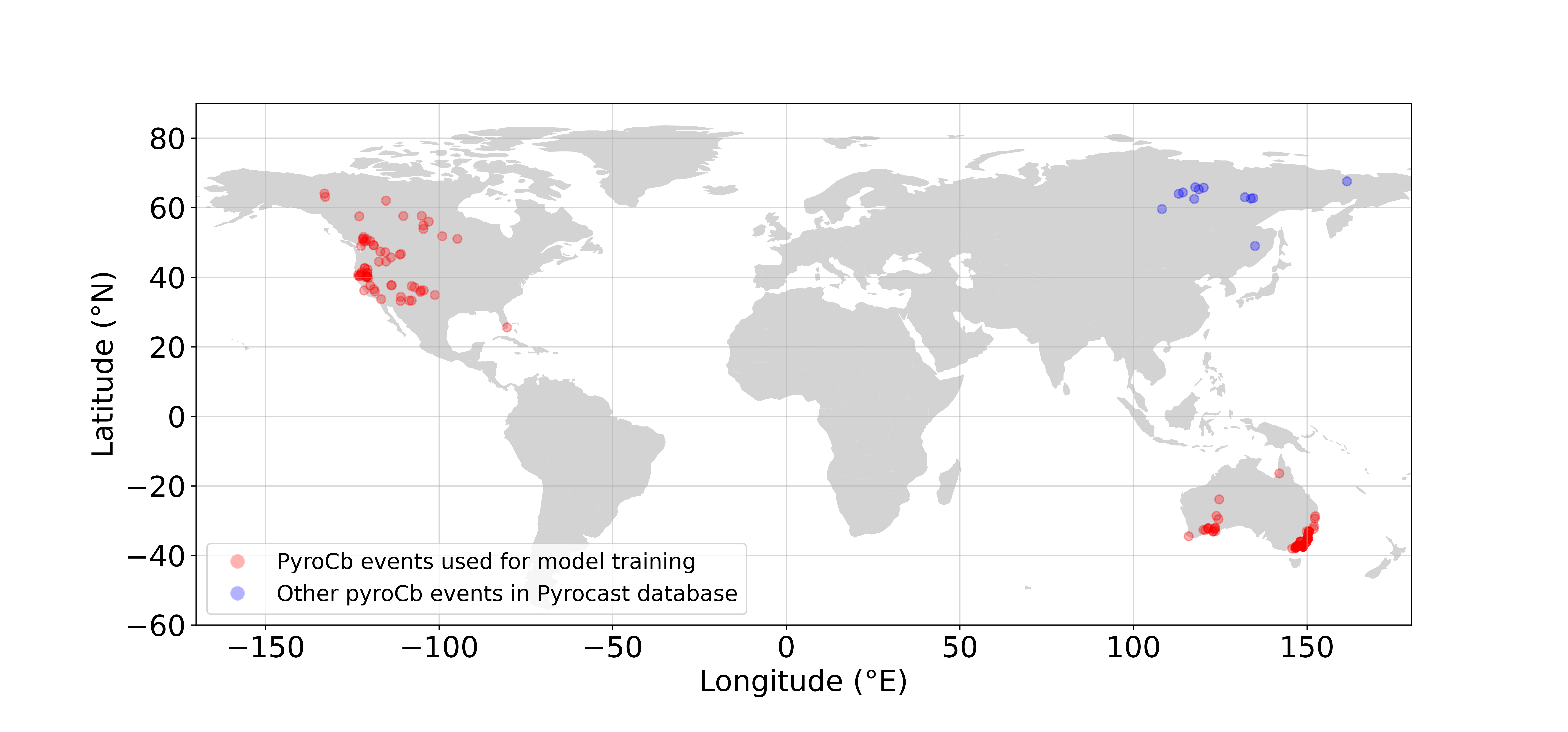}
        \caption{Spatial distribution of pyroCb events in the \textsc{Pyrocast} database.}  
        \label{fig:pyrocb_map}
\end{figure}

\begin{table}[ht]
\small
\caption{Geostationary satellite and ERA5 fuel and meteorological variables with description and motivation for selecting as pyroCb predictors.}
\label{intext_variabletable}
\centering
\begin{tabular}{p{.15\linewidth} p{.35\linewidth} p{.4\linewidth}}
\toprule
Variable & Description & Sensitive to \\
\midrule
\textit{ch}$1$ & $\phantom{\{0}0.47$\,\textmu m & smoke, haze\\
\textit{ch}$2$& $\phantom{\{0}0.64$\,\textmu m & terrain type\\
\textit{ch}$3$ & $\phantom{\{0}0.86$\,\textmu m & vegetation\\
\textit{ch}$4$ & $\phantom{\{0}3.9$\,\textmu m & thermal emissions \& cloud ice crystal size \\
\textit{ch}\{$5$,$6$\} & $\{11.2,13.3\}$\,\textmu m & thermal emissions \& cloud opacity \\
\midrule
\{\textit{u},\textit{v}\} & \{\textit{u},\textit{v}\} comp. of wind at $250$\,hPa & upper-lvl dynamics influencing rising motion \\
\{\textit{u},\textit{v}\}$10$ & $10$\,m \{\textit{u},\textit{v}\} component of wind & change in fire intensity and spread \\
\textit{fg}$10$ & $10$\,m  gusts since prev. post-processing & (same as above)\\
\midrule
\textit{blh} & boundary layer height & height of turbulent air at the surface \\
\midrule
\textit{cape} & convective available potential energy & energy for air to ascend into atmosphere\\
\textit{cin} & convective inhibition & energy that will prevent air from rising\\
\textit{z} & geopotential & energy needed for air to ascend into atmosphere as a function of altitude\\
\midrule
\{\textit{slhf}, \textit{sshf}\} & surface \{\text{latent}, \text{sensible}\} heat flux & heat released or absorbed \{\text{from}, \text{neglecting}\} phase changes\\
\midrule
\textit{w} & surface vertical velocity & ascent speed of the plume from the wildfire\\
\midrule
\textit{cv}\{\textit{h},\textit{l}\} & fraction of \{\text{high}, \text{low}\} vegetation & available fuel for the wildfire \\
\textit{type}\{\textit{H},\textit{L}\} & type of \{high, low\} vegetation & (same as above)  \\
\midrule
\textit{r}\{$650$,$750$,$850$\} & rel. humidity at \{$650$,$750$,$850$\}\,hPa & condensation of vapour into clouds \\
\bottomrule
\end{tabular}
\end{table}

\begin{figure}[h]
    \centering
        \includegraphics[scale=0.4]{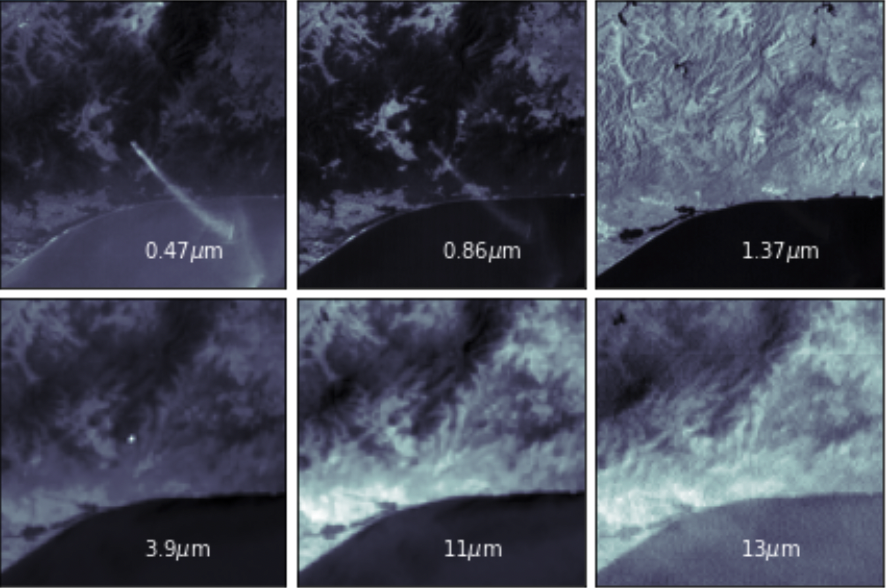}
        \caption{Example of geostationary images over the six wavelength channels a wildfire event in Timbarra, Australia (January 2019).}
        \label{fig:geo}
\end{figure}

\begin{figure}[h]
    \centering
        \includegraphics[scale=0.25]{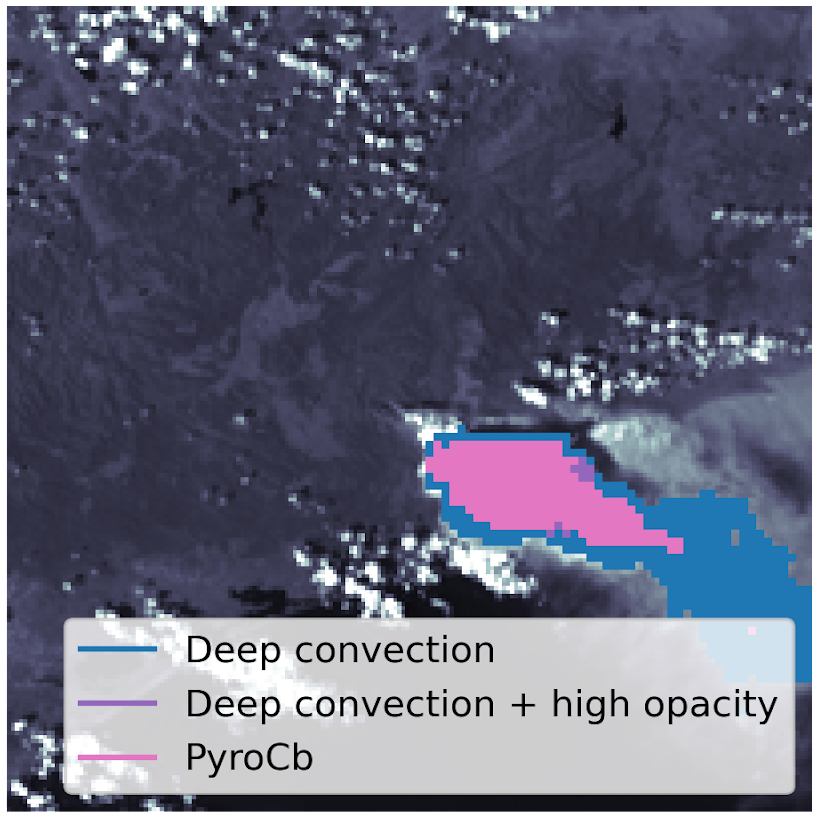}
        \caption{Example of NRL label masks overlaid onto geostationary satellite image (averaged over the $0.47$\,\textmu m, $0.64$\,\textmu m and $0.86$\,\textmu m channels) of a wildfire event in Timbarra, Australia (January 2019).}
        \label{fig:nrl}
\end{figure}

\begin{figure}[h]
    \centering
        \includegraphics[scale=0.8]{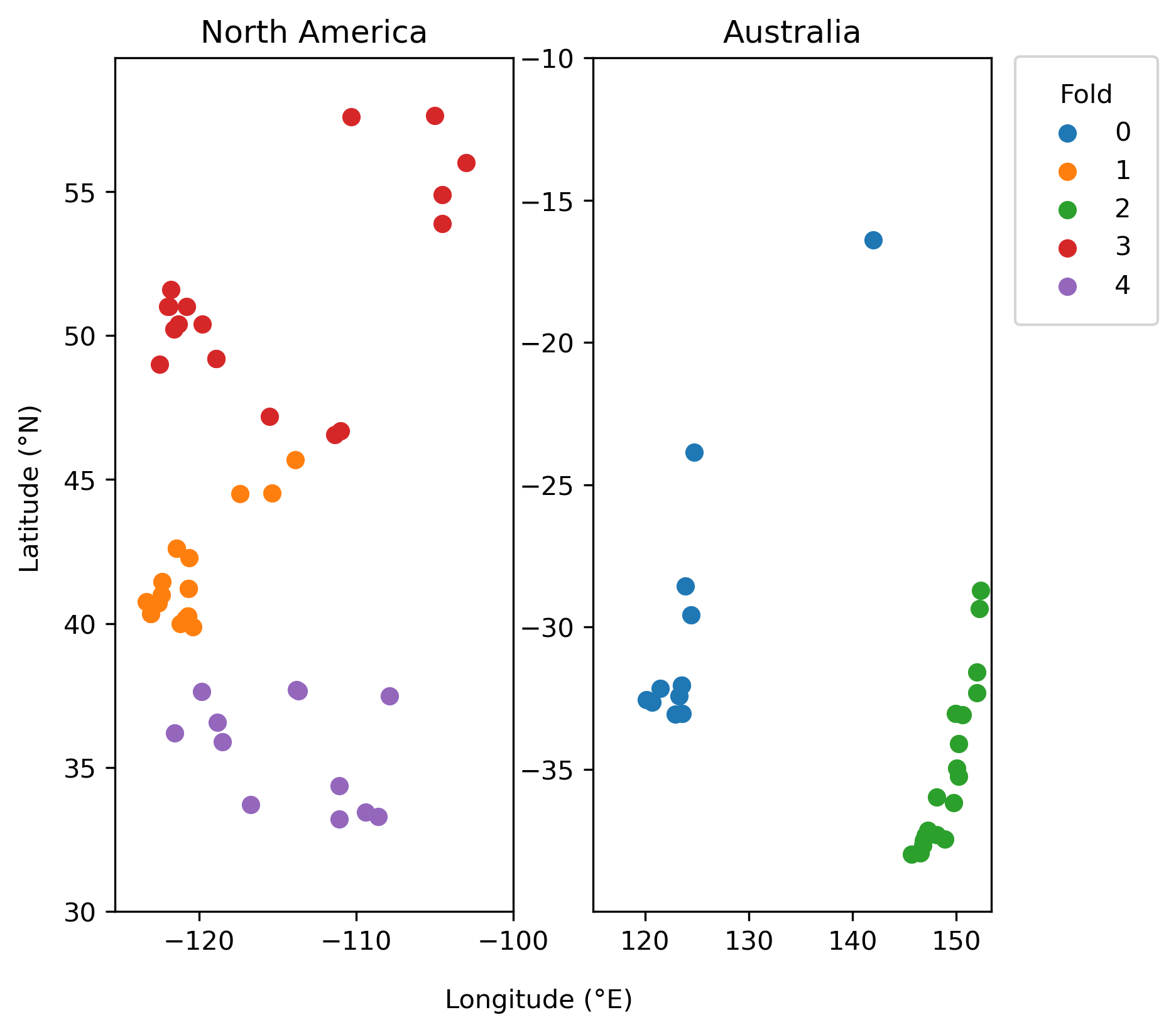}
        \caption{Clustered cross-validation folds.}
\end{figure}

\begin{table}[h] 
\caption{Average validation AUC across five clustered folds with standard deviations. These results are indicative model performance when generalising to an unseen region such as Europe.}
\label{auc_avg_clustered}
\centering
\small
\begin{tabular}{llccccc}
\toprule
 & & gs & w3 & gs+w3 & w19 & gs+w19\\

\midrule
Detection & RF & $0.94 \pm 0.02$ & $0.76 \pm 0.03$ & $0.93 \pm 0.03$ & $0.93 \pm 0.04$ & $0.93 \pm 0.04$\\ 
& CNN & $0.95 \pm 0.01$ & $0.65 \pm 0.07$ & $0.87 \pm 0.16$ &  & \\
& AE-CNN & $\mathbf{0.97 \pm 0.01}$ & $0.67 \pm 0.14$ & $0.91 \pm 0.01$ & & \\

\midrule
Forecast & RF & $0.70 \pm 0.08$ & $0.75 \pm 0.05$ & $0.76 \pm 0.05$ & $0.79 \pm 0.07$ & $\mathbf{0.80 \pm 0.06}$ \\
with oracle & CNN & $0.52 \pm 0.03$ & $0.65 \pm 0.06$ & $0.63 \pm 0.08$ &  & \\
 & AE-CNN & $0.58 \pm 0.05$ & $0.59 \pm 0.05$ & $0.68 \pm 0.06$ & & \\

\midrule
 Forecast & RF & $0.70 \pm 0.08$ & $0.70 \pm 0.06$ & $0.73 \pm 0.08$ & $0.74 \pm 0.05$ & $\mathbf{0.77 \pm 0.06}$\\
 & CNN & $0.52 \pm 0.03$ & $0.63 \pm 0.08$ & $0.64 \pm 0.09$ &  & \\ 
 & AE-CNN & $0.58 \pm 0.05$ & $0.66 \pm 0.06$ & $0.68 \pm 0.09$ &  & \\
\bottomrule
\end{tabular}
\end{table}

\end{document}